\documentclass[twocolumn,english]{revtex4-1}
\usepackage[T1]{fontenc}
\usepackage{color}
\usepackage{bm}
\usepackage{amsmath}
\usepackage{amssymb}
\usepackage{graphicx}

\makeatletter
\usepackage{bm}
\usepackage{subfigure}
\usepackage{multirow}
\usepackage{colortbl}
\usepackage{soul}
\usepackage{footmisc}

\usepackage{pdfcomment}
\usepackage{units}

\def\bbm[#1]{\mbox{\boldmath $#1$}}

\newcommand{\be}{\begin{equation}}
\newcommand{\ee}{\end{equation}}
\newcommand{\ba}{\begin{array}}
\newcommand{\ea}{\end{array}}
\newcommand{\bqa}{\begin{eqnarray}}
\newcommand{\eqa}{\end{eqnarray}}

\usepackage{babel}

\makeatother

\usepackage{babel}
\begin{document}

\title{Exceptional Points of Degeneracy and Branch Points for Transmission-Line
Problems \textendash{} Linear Algebra and Bifurcation Theory Perspectives}

\author{George W. Hanson}
\email{george@uwm.edu}

\selectlanguage{english}%

\address{Department of Electrical Engineering and Computer Science, University
of Wisconsin-Milwaukee, 3200 N. Cramer St., Milwaukee, Wisconsin 53211,
USA}

\author{Alexander B. Yakovlev}
\email{yakovlev@olemiss.edu }

\selectlanguage{english}%

\address{Department of Electrical Engineering, The University of Mississippi,
University, Mississippi 38677, USA}

\author{Mohamed Othman}
\email{mothman@uci.edu}

\selectlanguage{english}%

\address{Department of Electrical Engineering and Computer Science, University
of California-Irvine, Irvine, California 92697-2625, USA}

\author{Filippo Capolino }
\email{f.capolino@uci.edu }

\selectlanguage{english}%

\address{Department of Electrical Engineering and Computer Science, University
of California-Irvine, Irvine, California 92697-2625, USA}

\date{\today}
\begin{abstract}
We demonstrate several new aspects of exceptional points of degeneracy
(EPD) pertaining to propagation in two uniform coupled transmission-line
structures. We describe an EPD using two different approaches \textbf{\textendash{}}
by solving an eigenvalue problem based on the system matrix, and as
a singular point from bifurcation theory, and the link between these
two disparate viewpoints. Cast as an eigenvalue problem, we show that
eigenvalue degeneracies are always coincident with eigenvector degeneracies,
so that all eigenvalue degeneracies are implicitly EPDs in two uniform
coupled transmission lines. Furthermore, we discuss in some detail
the fact that EPDs define branch points (BPs) in the complex-frequency
plane; we provide simple formulas for these points, and show that
parity-time (PT) symmetry leads to real-valued EPDs occurring on the
real-frequency axis. We discuss the connection of the linear algebra
approach to previous waveguide analysis based on singular points from
bifurcation theory, which provides a complementary viewpoint of EPD
phenomena, showing that EPDs are singular points of the dispersion
function associated with the fold bifurcation. This provides an important
connection of various modal interaction phenomena known in guided-wave
structures with recent interesting effects observed in quantum mechanics,
photonics, and metamaterials systems described in terms of the EPD
formalism.
\end{abstract}
\maketitle

\section{Introduction}

When propagation in a coupled-waveguide system is described in terms
of a system matrix, exceptional points of degeneracy are points in
the parameter space of such a system at which simultaneous eigenvalue
and eigenvector degeneracies occur \cite{Heiss}. Interest in EPDs
has recently risen due to Parity-Time (PT) symmetric systems, wherein
non-Hermitian Hamiltonians can nevertheless exhibit real spectra,
representing physical observables. PT-symmetry has led to a range
of interesting phenomena in quantum mechanics and photonic systems
\cite{Bender1,Bender2,Ruter,Hodaei,Othman1,Hassan}, and in metamaterials
research \cite{Sounas,Monticone1,Monticone2,Chen1}, with applications
to cloaking, negative refraction, imaging, field transformation, and
sensing, among others. In a system whose evolution is described with
a system matrix, EPDs are associated with a Jordan block, corresponding
to a deficient (incomplete) set of eigenfunctions, and algebraically
growing solutions of generalized (associated) eigenvectors at the
EPD. Moreover, in the vicinity of EPDs, by virtue of small detuning,
eigenvalues exhibit unconventional perturbations following a fractional
power-law expansion in the perturbation parameters \cite{kato}.

It is important to point out that EPDs are manifest in the parameter
space of a system's eigenstates' temporal evolution (e.g., such as
certain coupled resonators with loss and gain), or of a system's eigenstates'
spatial evolution. This latter case represents the evolution of eigenwaves
in a given spatial direction, such as in a multimode waveguide with
prescribed loss and gain, which is investigated in this paper, where
the multimode waveguide is a pair of uniform coupled transmission
lines. Some of the earliest examples of EPDs have been also observed
in structures with spatial periodicity which are explored, for instance,
in \cite{FV1,FV2,OC,VS}, such as those exhibiting degenerate band
edges or stationary inflection points. Although EPDs are usually viewed
from a linear algebra standpoint, and are associated with systems
described by matrices with Jordan blocks \cite{FV1,Heiss}, it has
been observed that they also represent points in configuration space
where multiple branches of spectra connect, and are linked to branch
points in the space of control variables \cite{Hern,Hern2}.

In this work, we consider a coupled uniform transmission-line system,
recently examined in \cite{EPD}, and demonstrate several new aspects
of EPDs in these systems. Specifically, we stress that for a coupled
uniform transmission line, eigenvalue degeneracies always result in
eigenvector degeneracies, such that all eigenvalue degeneracies represent
EPDs. We derive closed-form expressions for the branch-point singularities/EPDs
using bifurcation theory. We show and discuss in detail the connection
of EPDs with previous work on fold-point and branch-point singularities
in waveguiding systems \cite{A1,A2,A3,A4,A6,A7,A11,A12,A14} associated
with mode degeneracies and mode interactions, which provides a complementary
viewpoint for understanding EPDs.

\section{Coupled Transmission-Line Formulation\label{sec:Coupled-Transmission-Line}}

We consider two uniform coupled transmission lines (CTLs) as depicted
in Fig. 1.

\begin{figure}[!th]
\noindent \begin{centering}
\includegraphics[width=3in]{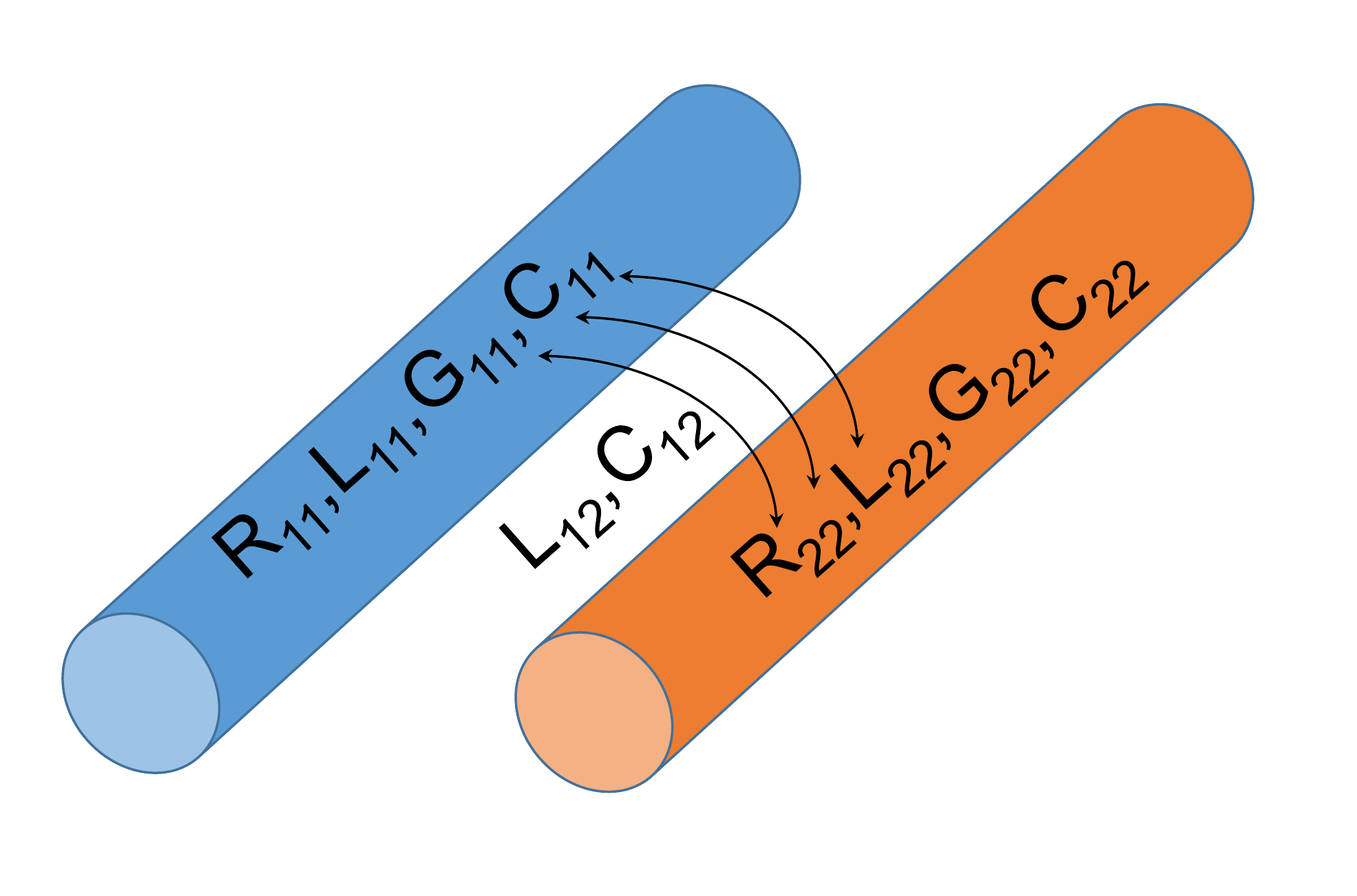} 
\par\end{centering}
\caption{Two coupled transmission lines with mutual capacitive and inductive
coupling, invariant along $z$. They exhibit EPDs under certain conditions
described in the paper.}
\label{fig1} 
\end{figure}

We refer to the formulation given in \cite{EPD} for the analysis
of eigenwaves propagating along the $z$-direction in a CTL (the $e^{i\omega t}$
time-harmonic evolution is implicitly assumed). Here we summarize
the mathematical steps carried out to obtain the eigenwaves supported
by such a guiding system. The CTL equations for a two-line network
consisting of uniform transmission lines are given by the telegraphers
equations \cite{Paul,MM} 
\begin{equation}
\frac{d\mathbf{V}\left(z\right)}{dz}=-\mathbf{\underline{\underline{\mathbf{Z}}}\ I}\left(z\right),\ \ \frac{d\mathbf{I}\left(z\right)}{dz}=-\mathbf{\underline{\underline{\mathbf{Y}}}\ V}\left(z\right)\label{GE}
\end{equation}
where the voltage and current are 2-dimensional vectors, $\mathbf{V}(z)=\left[V_{1}\left(z\right)\ \ V_{2}\left(z\right)\right]^{\text{T}}$
and $\mathbf{I}(z)=\left[I_{1}\left(z\right)\ \ I_{2}\left(z\right)\right]^{\text{T}}$,
whereas $\underline{\underline{\mathbf{Z}}}$ and $\underline{\underline{\mathbf{Y}}}$
are $2\times2$ matrices, 
\begin{equation}
\underline{\underline{\mathbf{Z}}}\left(\omega\right)=\left[\begin{array}{cc}
Z_{11} & Z_{12}\\
Z_{21} & Z_{22}
\end{array}\right],\ \ \underline{\underline{\mathbf{Y}}}\left(\omega\right)=\left[\begin{array}{cc}
Y_{11} & Y_{12}\\
Y_{21} & Y_{22}
\end{array}\right],
\end{equation}
 where the off-diagonal elements represent coupling between the two
transmission lines. Furthermore, the per-unit-length series impedance
and shunt admittance matrices are given by $\underline{\underline{\mathbf{Z}}}=i\omega\underline{\underline{\mathbf{L}}}+\underline{\underline{\mathbf{R}}}$
and $\underline{\underline{\mathbf{Y}}}=i\omega\mathbf{\underline{\underline{C}}}+\mathbf{\underline{\underline{G}}}$,
where $\mathbf{\underline{\underline{\mathbf{R}}},\underline{\underline{G}},\underline{\underline{\mathbf{L}}},\mathbf{}}$and
$\underline{\underline{\mathbf{C}}}$ are matrices of the per-unit-length
distributed CTL parameters, assumed nondispersive for simplicity.
The matrices $\underline{\underline{\mathbf{L}}}$ and $\mathbf{\mathbf{\underline{\underline{C}}}}$
are positive definite and symmetric \cite{Paul,MM}, and the off-diagonal
entries of $\mathbf{\mathbf{\underline{\underline{C}}}}$ and $\underline{\mathbf{\underline{G}}}$
are negative. In general $\underline{\underline{\mathbf{R}}}$ and
$\underline{\mathbf{\underline{G}}}$ are positive definite if they
represent losses (no gain), and in the following they are assumed
to be diagonal for simplicity. In addition, note also that the per-unit-length
impedance and admittance matrices may possess cutoff capacitance and
inductance terms, respectively, as done in Ch. 7 in \cite{H1961},
and also in \cite{OTper1} to model waveguide cutoff. Since we do
not investigate cutoff related degeneracies, we simply ignore these
terms in the CTL formulations above.

\subsection{EPD from a Linear Algebra Perspective}

Decoupling (\ref{GE}), we obtain two second-order wave equations
for the voltage and current vectors 
\begin{equation}
\frac{d^{2}\mathbf{V}\left(z\right)}{dz^{2}}=\mathbf{\underline{\underline{\mathbf{Z}}}\,\underline{\underline{\mathbf{Y}}}\ V}\left(z\right),\ \ \frac{d^{2}\mathbf{I}\left(z\right)}{dz^{2}}=\mathbf{\underline{\underline{\mathbf{Y}}}\,\underline{\underline{\mathbf{Z}}}\ I}\left(z\right).\label{2d}
\end{equation}
The two systems lead to the same wavenumber solutions though in general,
$\underline{\underline{\mathbf{Z}}}$ and $\underline{\underline{\mathbf{Y}}}$
do not necessarily commute; one common exception is for lossless lines
in a homogeneous environment characterized by $\mu,\varepsilon$,
in which case $\underline{\underline{\mathbf{Z}}}\,\underline{\underline{\mathbf{Y}}}=-\omega^{2}\mu\varepsilon\mathbf{\underline{\underline{1}}}$,
where $\underline{\underline{\mathbf{1}}}$ is the $2\times2$ identity
matrix. Alternatively, one may form a four-dimensional state vector
$\mathbf{\Psi}\left(z\right)=\left[V_{1}\left(z\right)\ \ V_{2}\left(z\right)\ \ I_{1}\left(z\right)\ \ I_{2}\left(z\right)\right]^{\text{T}}$,
leading to 
\begin{equation}
\frac{d}{dz}\mathbf{\Psi}\left(z\right)=-i\mathbf{\underline{M}}\left(\omega\right)\mathbf{\Psi}\left(z\right)\label{evolEq}
\end{equation}
where 
\begin{equation}
\mathbf{\underline{M}}\left(\omega\right)=\left[\begin{array}{cc}
\mathbf{\underline{\underline{0}}} & -i\underline{\underline{\mathbf{Z}}}\\
-i\underline{\underline{\mathbf{Y}}} & \underline{\underline{\mathbf{0}}}
\end{array}\right].\label{M}
\end{equation}
Assuming that the transmission line is invariant along $z$, the homogeneous
solutions to (\ref{2d}) and (\ref{evolEq}) are found to be in the
form $\mathbf{\Psi}\left(z\right)\propto e^{-ikz}$ with $k$ being
the wavenumber. As such, (\ref{2d}) and (\ref{evolEq}) become 
\begin{align}
-\left(\mathbf{\underline{\underline{\mathbf{Z}}}\,\underline{\underline{\mathbf{Y}}}}\right)\left(\omega\right)\mathbf{\ V}\left(z\right) & =k^{2}\mathbf{V}\left(z\right),\label{evp}\\
-\left(\underline{\underline{\mathbf{Y}}}\,\underline{\underline{\mathbf{Z}}}\right)\left(\omega\right)\mathbf{\ I}\left(z\right) & =k^{2}\mathbf{I}\left(z\right),\nonumber \\
\ \ \mathbf{\underline{M}}\left(\omega\right)\mathbf{\Psi}\left(z\right) & =k\mathbf{\Psi}\left(z\right).\nonumber 
\end{align}
Note that the first two equations in (\ref{evp}) have two eigenvalues
$k^{2}$ (and both signs of $k$ are possible), whereas the third
equation in (\ref{evp}) has four eigenvalues $k$. All three eigenvalue
problems lead to the same four eigenvalues, and encompass the same
physics, which is thoroughly explained in \cite{EPD}. Here, we wish
to make several new observations about these eigenproblems from two
different but complementary perspectives, which opens up new ways
for utilizing such EPDs and conceiving new operational principles
for a variety of microwave devices. For simplicity, we assume reciprocity,
i.e., $Y_{21}=Y_{12}$ and $Z_{21}=Z_{12}$. 

We denote the algebraic multiplicity for eigenvalues $\lambda$ (i.e.,
the order of the eigenvalue degeneracy) as $m(\lambda)$. The geometric
multiplicity of the eigenvalue (the span of the eigenvector space
associated with the eigenvalue) is denoted as $l(\lambda)$. We make
the following observations related to EPDs: 
\begin{enumerate}
\item For the systems of CTLs considered above, when an EPD occurs one has
$m\left(\lambda\right)>l\left(\lambda\right)$, i.e., all degenerate
eigenvalues have a deficient eigenspace, and the matrices $\mathbf{\underline{M},\underline{\underline{\mathbf{Z}}}\,\underline{\underline{\mathbf{Y}}}},\underline{\underline{\mathbf{Y}}}\,\underline{\underline{\mathbf{Z}}}$
\textit{cannot} be diagonalized (except for the trivial degeneracy
at $k$=0 and in uncoupled lines). In particular, for the two uniform
CTLs considered here, EPDs are associated with $l\left(\lambda\right)=2$,
and $m\left(\lambda\right)=1$. 
\item EPDs imply the presence of square-root branch points in the complex-frequency
plane. As such, these complex-frequency plane singularities are generally
unavailable for monochromatic problems, but may be accessed in certain
pulse shaping scenarios \cite{AA,CB,CB2}. 
\item The analysis of EPD from a linear algebra perspective can analogously
be studied as fold singularities of mappings in bifurcation theory. 
\item PT-symmetric conditions lead to EPDs on the real-frequency axis, and,
thus, to physically observable phenomena in monochromatic problems. 
\end{enumerate}
\bigskip{}

In the following, we examine the aforementioned statements and provide
analytical expressions for the eigenvalues and eigenvectors to reveal
the origin of EPDs and their relation to eigenvalue and eigenvector
degeneracies and branch points. In Section \ref{subsec:EPD-from-a}
we examine EPDs from a different prospective, that of bifurcation
theory.

We first consider the $2\,\textrm{\ensuremath{\times}}\,2$ eigenvalue
problem in (\ref{evp}); $-\mathbf{\underline{\underline{\mathbf{Z}}}\,\underline{\underline{\mathbf{Y}}}}$
having eigenvalues $k_{1,2}^{2}$ and $regular$ voltage eigenvectors
$\mathbf{V}_{1,2}$, obtained analytically as 
\begin{equation}
k_{n}^{2}=\frac{1}{2}\left(-T+(-1)^{n}D\right),\ \ \mathbf{V}_{n}=\left[\begin{array}{c}
-\frac{1}{2N_{1}}\left(N_{2}+(-1)^{n}D\right)\\
1
\end{array}\right]\label{eq:k1,2_V1,2}
\end{equation}
where $n=1,2$ , then $N_{1}=Y_{11}Z_{12}+Y_{12}Z_{22}$, $N_{2}=-Y_{11}Z_{11}+Y_{22}Z_{22}$
and 
\begin{equation}
D=\sqrt{T^{2}-4\text{det}\left(\mathbf{\underline{\underline{\mathbf{Z}}}\,\underline{\underline{\mathbf{Y}}}}\right)}.\label{D}
\end{equation}
The trace $T$ and determinant of $\underline{\underline{\mathbf{Z}}}\,\underline{\underline{\mathbf{Y}}}$
are given by 
\begin{align}
T & =\text{Tr}\left(\mathbf{\underline{\underline{\mathbf{Z}}}\,\underline{\underline{\mathbf{Y}}}}\right)=2Y_{12}Z_{12}+Y_{22}Z_{22}+Y_{11}Z_{11},\label{T}\\
\text{det}\left(\mathbf{\underline{\underline{\mathbf{Z}}}\,\underline{\underline{\mathbf{Y}}}}\right) & =\left(Y_{11}Y_{22}-Y_{12}^{2}\right)\left(Z_{11}Z_{22}-Z_{12}^{2}\right).\label{Trace}
\end{align}
For the $-\underline{\underline{\mathbf{Y}}}\,\underline{\underline{\mathbf{Z}}}$
formulation in (\ref{evp}), everything is analogous; the same eigenvalues
are obtained, and the regular current $\mathbf{I}_{1,2}$ eigenvectors
are retrieved using (\ref{eq:k1,2_V1,2}) by replacing $N_{1}\rightarrow Y_{22}Z_{12}+Y_{12}Z_{11}$.

It is obvious that, without considering the trivial eigenvalue degeneracy
at $k=0$, eigenvalue degeneracies occur when $D=0$, and, moreover,
from (\ref{eq:k1,2_V1,2}) it is clear that at this point eigenvectors
are also degenerate; $m\left(k^{2}\right)=2$ and $l\left(k^{2}\right)=1$
since $\mathbf{V}_{1}=\mathbf{V}_{2}$.

For the formulation in (\ref{evp}) involving the $4\,\textrm{\ensuremath{\times}}\,4$
matrix $\mathbf{\underline{M}}$, one finds the four eigenvalues and
$regular$ eigenvectors as 
\begin{align}
k_{n} & =\left(\pm\right)\frac{1}{\sqrt{2}}\sqrt{-T+\nu_{n}D},\ \ \label{eq:1psi}\\
\mathbf{\Psi}_{n} & \mathbf{=}\left[\begin{array}{c}
\left(\pm\right)i\frac{\sqrt{-T+\nu_{n}D}}{\sqrt{2}}\frac{-N_{2}-\nu_{n}D}{N_{3}-\nu_{n}YD}\\
\left(\pm\right)i\frac{2\sqrt{-T+\nu_{n}D}}{\sqrt{2}}\frac{N_{1}}{N_{3}-\nu_{n}Y_{12}D}\\
\frac{\left(-N_{2}-\nu_{n}D\right)Y_{11}+2Y_{12}N_{1}}{N_{3}-\nu_{n}Y_{12}D}\\
1
\end{array}\right],\label{kPsi}
\end{align}
where the $+$ sign in front is for $n=1,2$ , the $-$ sign in front
is for $n=3,4$ , $\nu_{n}=(-1)^{n}$, $N_{3}=Y_{11}\left(Y_{12}Z_{11}+2Y_{22}Z_{12}\right)+Y_{12}Y_{22}Z_{22}$,
and again both eigenvalues and eigenvectors become simultaneously
degenerate when $D=0$, and $m_{\mathbf{}}\left(\pm k\right)=2>l\left(\pm k\right)=1$. 

Therefore, excepting the case of uncoupled identical lines \cite{FN}
and $k=0$, for all system descriptions in (\ref{evp}) eigenvector
degeneracies are simultaneous with eigenvalue degeneracies. Thus,
these simultaneous eigenvalue and eigenvector degeneracies are, by
definition, an EPD, where $k=\pm k_{e}$ with $k\equiv\sqrt{-T}/\sqrt{2}$.
Indeed at such points the matrices in (\ref{evp}) are deficient and
cannot be diagonalized because there are not enough eigenvectors to
form a complete basis. This proves Item 1 above. From the above analysis,
Item 2 is also demonstrated, since $D=D\left(\omega\right)$ clearly
represents a square-root type branch point in the complex-$\omega$
plane.

Conditions for EPDs were also presented in \cite{EPD}; here we briefly
comment on those and the connection with the condition $D=0$. In
\cite{EPD}, it was shown that the conditions 
\begin{align}
 & T=\text{Tr}\left(\mathbf{\underline{\underline{\mathbf{Z}}}\,\underline{\underline{\mathbf{Y}}}}\right)=-2k^{2},\label{tr}\\
 & \text{det}\left(\mathbf{\underline{\underline{\mathbf{Z}}}\,\underline{\underline{\mathbf{Y}}}}\right)=k^{4},\label{det3}
\end{align}
are necessary for an eigenvalue degeneracy (and so, in fact, are necessary
and sufficient for an EPD as described previously, excepting $k=0$
and uncoupled lines). These two conditions combined yield det$\left(\mathbf{\underline{\underline{\mathbf{Z}}}\,\underline{\underline{\mathbf{Y}}}}\right)=T^{2}/4$,
which is the condition under which $D=0$.

Furthermore, when, e.g., $\mathbf{\underline{M}}$ is similar to a
diagonal matrix (away from the EPD) it can be written in the form
\begin{equation}
\mathbf{\underline{M}}=\mathbf{\underline{U}}\,\mathbf{\underline{\Lambda}}\,\mathbf{\underline{U}}^{-1}\label{SimM}
\end{equation}

\noindent where $\mathbf{\underline{U}}\,$ is a $4\,\textrm{\ensuremath{\times}}\,4$
matrix representing the similarity transformation of $\mathbf{\underline{M}}$
that brings it to a diagonal form, and $\mathbf{\underline{\Lambda}}$
is a diagonal matrix whose diagonal entries are the eigenvalues $k_{n}$
in (\ref{eq:1psi}). It was shown in \cite{EPD} that the condition
$\mathrm{det}\left(\mathbf{\underline{U}}\right)=0$ provides necessary
and sufficient conditions for an eigenvector degeneracy (at which
point the regular eigenvectors must be augmented with associated eigenvectors,
and, rather than a diagonal form, the simplest matrix representation
is given by the Jordan canonical form \cite{LT}). Forming\textit{\ }
\begin{equation}
\det\left(\mathbf{\underline{U}}\right)=-16\frac{Y_{11}}{N_{2}^{3}}D^{2}\left(Y_{12}Z_{22}+Y_{11}Z_{12}\right)\sqrt{T^{2}-D^{2}}=0\label{U}
\end{equation}
it is observed that $\mathrm{det}\left(\mathbf{\underline{U}}\right)=0$
occurs when $D=0$ (or when $Y_{12}Z_{22}+Y_{11}Z_{12}=0$, which
seems to not be of practical interest, and note that $D=T$ cannot
be true since, using (\ref{D}), it would hold only if det$\left(\mathbf{\underline{\underline{\mathbf{Z}}}\,\underline{\underline{\mathbf{Y}}}}\right)=0$,
which is not true). Alternatively, assuming a similarity transformation
analogous to that in (\ref{SimM}) but that diagonalizes the 2 $\times$
2 matrix $-\mathbf{\underline{\underline{\mathbf{Z}}}\,\underline{\underline{\mathbf{Y}}}}$,
\begin{equation}
\det\left(\mathbf{\underline{\underline{\mathbf{U}}}}\right)=-\frac{D}{Y_{11}Z_{12}+Y_{12}Z_{22}}
\end{equation}
which again occurs at $D=0$. Therefore, the previously stated conditions
in \cite{EPD} are, for uniform CTLs modeled by nondispersive $\mathbf{\underline{\underline{\mathbf{R}}},\underline{\underline{G}},\underline{\underline{\mathbf{L}}},\mathbf{}}$and
$\underline{\underline{\mathbf{C}}}$ parameters, alternative ways
of stating the $D=0$ EPD condition. 

\textbf{\textcolor{black}{Puiseux series}}\textcolor{black}{. In what
follows, it will be useful to cast the eigenvalue problems (\ref{evp})
in the form 
\begin{equation}
H\left(k,\omega\right)=\mathrm{det}\left(\mathbf{A}\left(\omega,\bm{\xi}\right)-k\mathbf{1}\right)=0\label{eq:H}
\end{equation}
where $\bm{\xi}$ is the vector of geometrical and material parameters
of the system, and $\mathbf{1}$ is the identity matrix. In particular,
in the following, all the partial derivatives in $\omega$ could be
substituted with partial derivatives in $\bm{\xi}$ and analogous
conclusions would be reached relative to the dispersion diagram $\left(k,\bm{\xi}\right)$
and associated BPs. In (\ref{eq:H}), the matrix $\mathbf{A}$ represents
either the 2$\times$2 system for which $\mathbf{\underline{\underline{\mathbf{A}}}=-\underline{\underline{\mathbf{Z}}}\,\underline{\underline{\mathbf{Y}}}}$,
or $\underline{\underline{\mathbf{A}}}=-\mathbf{\underline{\underline{\mathbf{Y}}}\,\underline{\underline{\mathbf{Z}}}}$
(in which case the eigenvalue is $k^{2}$ rather than $k$) or the
4$\times$4 system $\mathbf{\underline{\mathbf{A}}=\underline{M}}$.
In the following we suppress the dependence on $\bm{\xi}$. The condition
(\ref{eq:H}) leads to }
\begin{equation}
k^{4}+k^{2}\text{Tr}\left(\underline{\underline{\mathbf{A}}}\right)+\text{det}\left(\underline{\underline{\mathbf{A}}}\right)=0,\label{HE0}
\end{equation}
which is also given in \cite{EPD}. \textcolor{black}{Denoting derivatives
as 
\begin{equation}
H_{\varsigma}^{(m)}\left(k_{e},\omega_{e}\right)=\left.\frac{\partial^{(m)}H(k,\omega)}{\partial\varsigma^{m}}\right|_{(k_{e},\omega_{e})},
\end{equation}
for $\varsigma=k,\omega$, an }\textit{\textcolor{black}{m}}\textcolor{black}{th-order
eigenvalue degeneracy (i.e., an }\textit{\textcolor{black}{m}}\textcolor{black}{th-order
root of $H\left(k,\omega\right)$) will satisfy 
\begin{equation}
H(k_{e},\omega_{e})=H_{k}^{\prime}(k_{e},\omega_{e})=...=H_{k}^{(m-1)}(k_{e},\omega_{e})=0,\label{eq:H1}
\end{equation}
\begin{equation}
H_{k}^{(m)}\left(k_{e},\omega_{e}\right)\neq0,\label{H2}
\end{equation}
}

\noindent \textcolor{black}{where $k_{e}$ is the degenerate wavenumber
and $\omega_{e}$ is the frequency at which the wavenumbers become
degenerate. }For a second-order EPD, the condition $H_{k}^{\prime}(k,\omega)=0$
is 
\begin{equation}
k\left(T+2k^{2}\right)=0,
\end{equation}
which is equivalent to the trace condition (\ref{tr}) for $k\neq0$,
and leads to $k=\pm\sqrt{-T}/\sqrt{2}$, consistent with the general
eigenvalue at the EPD.\textcolor{black}{{} As described briefly in \cite{EPD}
but of more direct importance here, the eigenvalues of the CTL at
such a degeneracy can be written as a convergent Puiseux series \cite{AM,kato}
\begin{equation}
k_{n}(\omega)=k_{e}+\alpha_{1}\zeta^{n}(\omega-\omega_{e})^{\frac{1}{m}}+{\displaystyle \sum_{p=2}^{\infty}\alpha_{p}}(\zeta^{n}(\omega-\omega_{e})^{\frac{1}{m}})^{p}\label{eq:PS}
\end{equation}
for $n=0,1,2,...,m-1$, where $\zeta=e^{i\frac{2\pi}{m}}$. The first-order
coefficient is given by 
\begin{equation}
\alpha_{1}=\left(-\frac{H'_{\omega}\left(k_{e},\omega_{e}\right)}{\frac{1}{m!}H{}_{k}^{(m)}\left(k_{e},\omega_{e}\right)}\right)^{\frac{1}{m}}.\label{eq:alpha_1}
\end{equation}
The Puiseux series is a direct consequence of the Jordan Block form
(see for example page 65 in \cite{kato}) hence it is always relevant
in systems that exhibit an EPD to describe the eigenvalue perturbation
away from the EPD. Applying the fractional power expansion (\ref{eq:PS})
to the 2nd order EPD in the uniform CTL above, and ignoring expansion
terms with order equal or higher than $\omega-\omega_{e}$, one arrives
at}

\noindent \textcolor{black}{
\begin{equation}
k(\omega)\simeq k_{e}\pm\alpha_{1}\sqrt{(\omega-\omega_{e})}+O(\omega-\omega_{e}).\label{eq:EPD_first_terms}
\end{equation}
The first two terms in (\ref{eq:EPD_first_terms}) show the occurrence
of the branch-point singularity in the complex-frequency plane, resulting
from the square-root function. Associated with this series is the
condition \cite{AM} 
\begin{equation}
H'_{\omega}\left(k_{e},\omega_{e}\right)\neq0,\label{H3}
\end{equation}
and so the first-order coefficient $\alpha_{1}$ is nonzero. An important
aspect of the Puiseux series is that it provides the characteristic
form of the solution in the vicinity of the EPD, as shown later in
relation to Fig. 2. Regarding Statement 3, the conditions (\ref{eq:H1})-(\ref{H2}),
and (\ref{H3}) will be reconsidered in Section \ref{subsec:EPD-from-a}
from the viewpoint of singularity and bifurcation theory.}

\textbf{Jordan Block and Generalized Eigenvectors. }At an EPD in a
uniform 2-CTL, the eigenvalue degeneracy corresponds to an eigenvector
degeneracy as we have previously discussed. This can be also shown
by noticing that when the eigenvalues of a $2\times2$ system matrix,
as in the first two systems in (\ref{evp}), are identical then it
is either proportional to an identity matrix (hence with two independent
eigenvectors) or otherwise it must be proportional to a $2\times2$
Jordan block (that exhibit the eigenvector degeneracy). For the $4\times4$
system matrix $\mathbf{\underline{M}}$ as in the third system in
(\ref{evp}) the situation is more involved. At an EPD the system
matrix $\mathbf{\underline{M}}$ is similar to a matrix containing
two Jordan blocks as

\begin{equation}
\mathbf{\underline{M}}=\mathbf{\underline{U}}\left[\begin{array}{cc}
\mathbf{\underline{\underline{\mathbf{J}}}_{+}} & \mathbf{\underline{\underline{0}}}\\
\mathbf{\underline{\underline{0}}} & \mathbf{\underline{\underline{\mathbf{J}}}_{-}}
\end{array}\right]\mathbf{\mathbf{\underline{U}}}^{-1},\,\,\,\,\,\,\,\,\,\,\,\,\,\,\mathbf{\underline{\underline{\mathbf{J}}}_{\pm}}=\left[\begin{array}{cc}
\pm k_{e} & 1\\
0 & \pm k_{e}
\end{array}\right]
\end{equation}
where $\mathbf{\underline{U}}$ is a 4$\times$4 matrix constituting
a similarity transformation and containing the generalized eigenvectors
of $\mathbf{\underline{M}}$ namely $\mathbf{\underline{S}}=\left[\Psi_{1}\,\,|\,\,\Psi_{1}^{g}\,\,|\,\,\Psi_{3}\,\,|\,\,\Psi_{3}^{g}\right]$
that are constructed through the Jordan chain procedure (\cite{EPD},
\cite{Meyer}, see also \cite{HN} for the differential operator case)

\begin{align}
\left(\mathbf{\underline{M}}-k_{e}\mathbf{\underline{1}}\right)\Psi_{1}\, & =0,\,\,\,\,\,\,\,\,\,\,\,\left(\mathbf{\underline{M}}-k_{e}\mathbf{\mathbf{\underline{1}}}\right)\Psi_{1}^{g}\,=\Psi_{1}\\
\left(\mathbf{\underline{M}}+k_{e}\mathbf{\mathbf{\underline{1}}}\right)\Psi_{3}\, & =0,\,\,\,\,\,\,\,\,\,\,\,\left(\mathbf{\underline{M}}+k_{e}\mathbf{\mathbf{\underline{1}}}\right)\Psi_{3}^{g}\,=\Psi_{3}\label{GeneralizedPsi}
\end{align}
with $\Psi_{1}$ and $\Psi_{1}^{g}$ being the regular and generalized
eigenvectors associated with the wavenumber $k_{e}$ at the second-order
EPD, and similarly $\Psi_{3}$ and $\Psi_{3}^{g}$ are the regular
and generalized eigenvectors associated with the wavenumber $-k_{e}$. 

We consider the general solution of (\ref{evolEq}) subject to an
initial condition at an arbitrary $z=z_{0}$ given by $\Psi(z_{0})=\Psi_{0}.$
Its general and unique solution is given by

\begin{align*}
\Psi(z) & =\exp(-i\mathbf{\underline{M}}z)\Psi_{0}\\
 & =\mathbf{\underline{U}}\left[\begin{array}{cc}
\exp(-i\mathbf{\underline{\underline{\mathbf{J}}}_{+}}z) & \underline{\mathbf{\underline{0}}}\\
\underline{\mathbf{\underline{0}}} & \exp(-i\mathbf{\underline{\underline{\mathbf{J}}}_{+}}z)
\end{array}\right]\mathbf{\underline{U}}^{-1}\Psi_{0},\\
 & =\mathbf{\mathbf{\underline{U}}}\left(\left[\begin{array}{cccc}
e^{-ik_{e}z} & -ize^{-ik_{e}z} & 0 & 0\\
0 & e^{-ik_{e}z} & 0 & 0\\
0 & 0 & e^{+ik_{e}z} & ize^{ik_{e}z}\\
0 & 0 & 0 & e^{+ik_{e}z}
\end{array}\right]z\right)\mathbf{\mathbf{\underline{U}}}^{-1}\Psi_{0}
\end{align*}
which provides growing solutions along $z$ as $\Psi(z)\propto ze^{-ik_{e}z}$
discussed in \cite{EPD}. 

\subsection{EPD from a Theory of Singular and Bifurcation Points Perspective\label{subsec:EPD-from-a}}

Here, we address Statement 3, and connect the previous analysis with
an entirely different method based on singularity and bifurcation
theory \cite{A9,A10}. We consider the implicit dispersion equation
(\ref{eq:H}), $H(k,\omega)=\mathrm{det}(\mathbf{A}(\omega)-k\mathbf{1})=0.$
Here, $H(k,\omega)$ is more generally understood as a mapping $\mathrm{\mathbb{C^{\mathrm{2}}\rightarrow C}}$,
$H(k,\omega)=z$. Obviously, the modal solutions of interest occur
for $z=0$, although viewing $H$ more generally as a mapping facilitates
the analysis below. For many waveguiding structures one must solve
$H(k,\omega)=0$ numerically, via a complex-plane root search, but
for the CTLs of interest here an explicit solution can be obtained,
$k_{n}(\omega)=\left(\pm\right)\frac{1}{\sqrt{2}}\sqrt{-T+\nu_{n}D},$
as given in (\ref{eq:1psi}).

\begin{figure}[!th]
\noindent \begin{centering}
\includegraphics[width=3in]{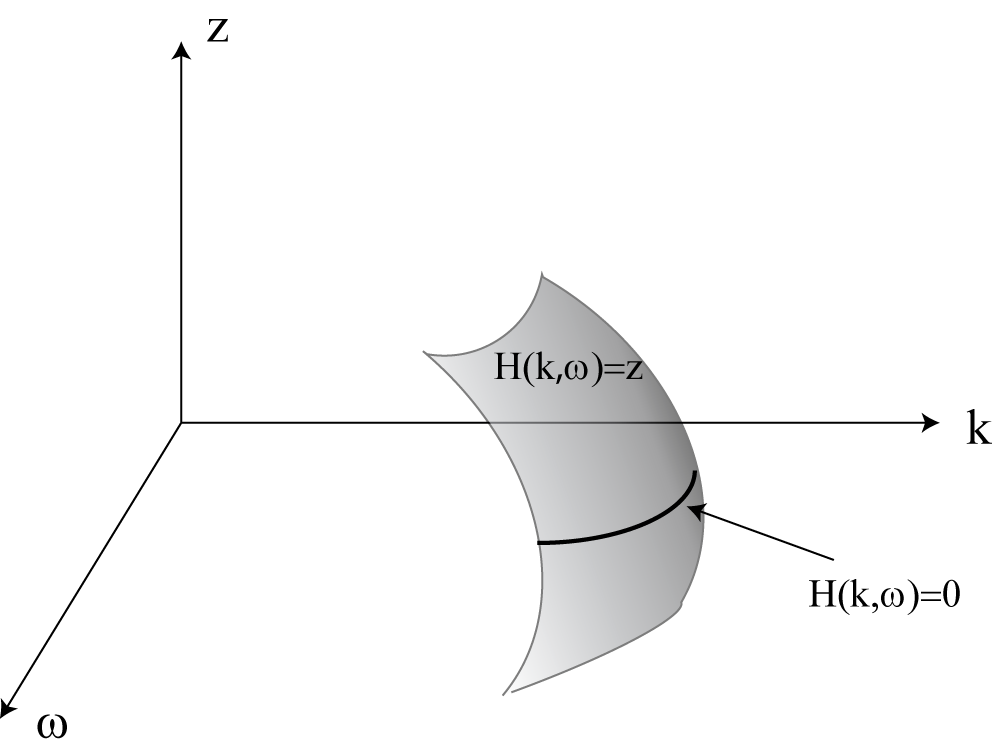} 
\par\end{centering}
\caption{Depiction of the surface defined by $H(k,\omega)=z$, for $(k,\omega,z)\in\mathbb{R}$.
The surface $H(k,\omega)=z$ may intersect the $(k,\omega)-$plane,
at $H(k,\omega)=0$, resulting in the curved line of intersection
shown that represents a standard dispersion diagram. If $H(k,\omega)=0$
does not have solutions for $(k,\omega)\in\mathbb{R}$, then solutions
can be be found in complex space. }
\label{fig1-1} 
\end{figure}

The mapping $H(k,\omega)=z$ defines a surface in $\mathbb{C\mathrm{^{2}}}$,
and for the simple case of $(k,\omega,z)\in\mathbb{R}$, this is depicted
in Fig. \ref{fig1-1}. The particular case of $H(k,\omega)=0$ defines
a curve (solid line in Fig. \ref{fig1-1}), which is the dispersion
curve of interest, and the smoothness of that curve at a given point
determines important modal properties. In particular, one can define
regular and singular points of the curve associated with certain modal
behavior \cite{A1,A2,A3,A4}. In the following we consider $k$ as
the unknown and $\omega$ as a distinguished parameter, although the
roles can also be reversed.

We first define a regular point on the curve $H(k,\omega)=0$ as a
point where $\partial H/\partial k\neq0$. At a regular point the
implicit function theorem \cite{A5} can be used to show that a unique
smooth curve $k=k(\omega)$ exists in the neighborhood of the point.
Except for a finite number of non-regular points (a set of measure
zero), all points of modal dispersion are regular points, wherein
the dispersion curve is smooth and single-valued. It is also worthwhile
to note that differentiation $d/d\omega$ of $H(k,\omega)=0$ leads
to, via the chain rule, 
\begin{equation}
\frac{dk}{d\omega}=-\frac{\partial H/\partial\omega}{\partial H/\partial k},\label{eq:T}
\end{equation}
and, therefore, at a regular point the tangent of $k(\omega)$ (related
to the group velocity) is well-defined. However, of particular interest
are the singular points \cite{A4} of the mapping $H$, which ultimately
lead to branch points in the complex-frequency plane \cite{A2,A6}.
The point $(k_{s},\omega_{s})$ is said to be a singular point of
the mapping $H$ if \cite[p. 2]{A9}

\begin{equation}
H(k_{s},\omega_{s})=H_{k}^{\prime}(k_{s},\omega_{s})=0.\label{Eq4-1}
\end{equation}

\begin{figure}[!th]
\noindent \begin{centering}
\includegraphics[scale=0.5]{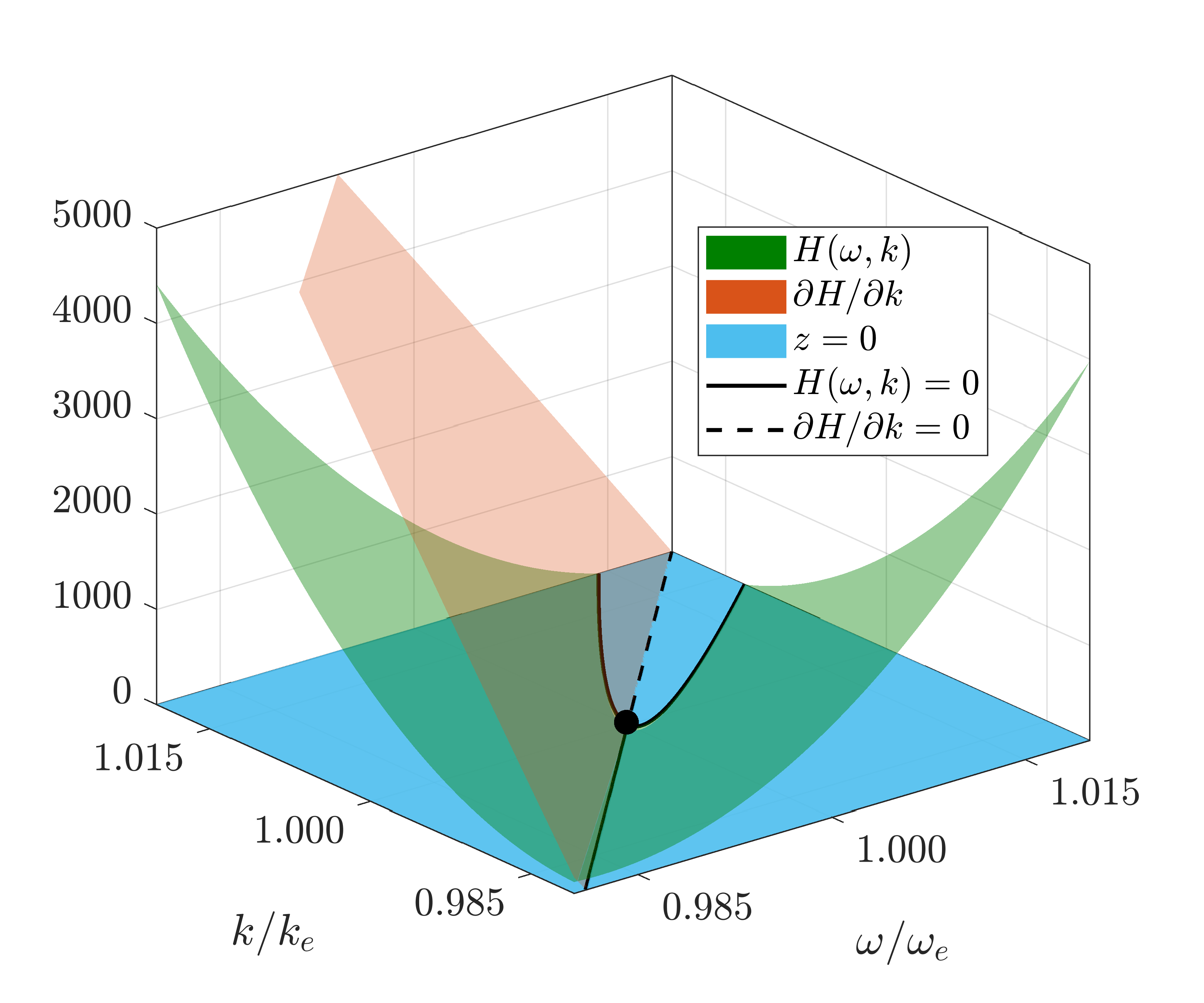} 
\par\end{centering}
\caption{The functions $H(k,\omega)$ (green), $H'_{k}(k,\omega)$ (pink),
and the zero plane (blue) vs. $k,\omega$ in the vicinity of the EPD
(solid dot). The 2D dispersion $H(k,\omega)=0$ is also shown (solid
black line). The black dashed line is $H_{k}^{\prime}(k,\omega)=0$.
Units of $H(k,\omega)$ (green) and $H'_{k}(k,\omega)$ (pink) are
in $m^{-4}$ and $m^{-3}$ , respectively.}
\label{fig1-2} 
\end{figure}

\noindent Obviously, in this case the tangent (\ref{eq:T}) is undefined.
In \cite[p. 45]{A9} it is shown that $H_{k}^{\prime}(k_{s},\omega_{s})=0$
is a necessary condition for the solution of $H(k_{s},\omega_{s})=0$
to be a bifurcation point (a point where the number of solutions changes).
For the two coupled transmission lines described above, Fig. \ref{fig1-2}
shows a plot of $H(k,\omega)$ in the vicinity of the EPD $(k_{s},\omega_{s})=(k_{e},\omega_{e})$
(the green, curved surface; numerical values of the CTL parameters
are the same as given in Section \ref{subsec:EDPs-as-branch}). The
intersection with the zero plane (solid blue) is clearly visible,
which forms the dispersion curve; the 2D dispersion is shown as the
black solid line (see also Fig. 4a in \cite{EPD}). A plot of the
function $H_{k}^{\prime}(k,\omega)$ is also shown in Fig. \ref{fig1-2}
(the slanted orange plane); units of $H(k,\omega)$ (green) and $H'_{k}(k,\omega)$
(pink) are in $m^{-4}$ and $m^{-3}$ , respectively. The intersection
of $H_{k}^{\prime}(k,\omega)$ with the $z=0$ plane forms the line
$H_{k}^{\prime}(k,\omega)=0$ shown in the figure with a black dashed
curve. The intersection of $H(k,\omega)$ and $H_{k}^{\prime}(k,\omega)$
on the $z=0$ plane is at the singular point (EPD) denoted by a black
solid circle (note that for $\omega<\omega_{e}$ the solid and dashed
curves seem to overlap. This is merely due to the scale of the plot;
the two lines actually only intersect at the EPD). For both $H$ and
$H_{k}^{\prime}$ the real part of the function is shown, as the imaginary
parts are negligible. 

In addition to the conditions (\ref{Eq4-1}), we defined a fold bifurcation
point (also know as a turning point, or limit point) when $H$ satisfies
(\ref{Eq4-1}) together with 

\begin{equation}
H_{kk}^{\prime\prime}(k_{s},\omega_{s})\neq0,\,\,\,H_{\omega}^{\prime}(k_{s},\omega_{s})\neq0.\label{Eq4-1a}
\end{equation}
The zero conditions (\ref{Eq4-1}) together with the nonzero condition
$H_{kk}^{\prime\prime}(k_{s},\omega_{s})\neq0$ indicates that the
degeneracy is of second-order, i.e., where two modal eigenvalues coalesce,
as given in (\ref{eq:H1})-(\ref{H2}). The\textcolor{black}{{} nonzero
condition $H_{\omega}^{\prime}(k_{s},\omega_{s})\neq0$ serves as
a sufficient condition for $\omega_{s}$ to be a BP in the complex
$\omega-$plane, as proved in \cite{A2} using the Weierstrass preparation
theorem. }In \cite{A1,A2,A3,A4,A7,A8} the importance of fold singular
points in modal interaction phenomena on guided-wave structures has
been addressed in connection with the fold bifurcation from bifurcation
theory \cite{A9,A10}. 

\textcolor{black}{Notably, the zero and non-zero conditions }(\ref{Eq4-1})-(\ref{Eq4-1a})\textcolor{black}{{}
are the same as (\ref{eq:H1})-(\ref{H2}), and (\ref{H3}) that arise
from linear algebra analysis. Thus, it can be concluded that the fold
singular point considered in, e.g., \cite{A1,A2,A3,A4,A7,A8} is in
fac}t an EPD which may reside generally in the complex plane $(k,\omega)\in\mathbb{C}^{2}$.
Therefore, in the following we denote $(k_{s},\omega_{s})$ as $(k_{e},\omega_{e})$.\textcolor{red}{{}
}An analogous treatment of EPDs using the conventional coupled-mode
theory \cite{CMT} is briefly outlined in the appendix.

\textbf{\textcolor{black}{Characteristic form. }}In the local neighborhood
of the fold point (FP)/EPD $(k_{e},\omega_{e})$ the qualitative behavior
of the mapping $H$ can be represented by the normal form \cite[p. 308-309]{A10}
, \cite[p. 196-198]{A9}, 
\begin{align}
(k-k_{e})^{2}+(\omega-\omega_{e})=0,\,\,\,\Delta & >0,\label{Eq7}\\
(k-k_{e})^{2}-(\omega-\omega_{e})=0,\,\,\,\Delta & <0\nonumber 
\end{align}
where $\Delta=H_{kk}^{\prime\prime}(k_{e},\omega_{e})H_{\omega}^{\prime}(k_{e},\omega_{e})$,
leading to the dispersion function 
\begin{align}
k(\omega) & =k_{e}\pm i\sqrt{\omega-\omega_{e}},\,\,\,\Delta>0,\label{Eq8}\\
k(\omega) & =k_{e}\pm\sqrt{\omega-\omega_{e}},\,\,\,\Delta<0.\nonumber 
\end{align}
For the case of $\Delta>0$ with $\omega<\omega_{e}$ two branching
solutions $\Re{(k(\omega))}$ of $(k-k_{e})^{2}+(\omega-\omega_{e})$
generate a parabola, and for $\omega>\omega_{e}$ two equal solutions
$\Re{(k(\omega))}$ exist as a straight line $k(\omega)=k_{e}$. This
corresponds to the characteristic intersection of a parabola and a
straight line that occurs at a point of fold bifurcation \cite{A9,A10},
as shown in Fig. \ref{fig1-2} (see also \cite{EPD}). When $\omega=\omega_{e}$
there is only one solution $(k_{e},\omega_{e})$ corresponding to
the fold point. Also, $\Im{(k(\omega))}$ for $\omega<\omega_{e}$
yields the solution $k(\omega)=0$, and for $\omega>\omega_{e}$ two
branching solutions form a parabola in the imaginary plane of $k(\omega)$.
A similar analysis can be applied to the case of $\Delta<0$. 

It should be noted that the conditions (\ref{Eq4-1}) and (\ref{Eq4-1a})
define both real and complex FPs/EPDs, however, the normal form (\ref{Eq7})
is applicable for real valued FPs, where $\Delta$ is real-valued.
Otherwise, the quantitative behavior of the local structure of the
function $H(k,\omega)$ in the vicinity of FP/EPD can be obtained
with a Taylor series expansion. Explicitly, the Taylor series in the
vicinity of the EPD can be written as 
\begin{align}
H\left(k,\omega\right)= & H\left(k_{e},\omega_{e}\right)+H_{k}^{\prime}\left(k-k_{e}\right)+H_{\omega}^{\prime}\left(\omega-\omega_{e}\right)\\
 & +\frac{1}{2}H_{kk}^{\prime\prime}\left(k-k_{e}\right)^{2}+H_{k\omega}^{\prime\prime}\left(k-k_{e}\right)\left(\omega-\omega_{e}\right)\nonumber \\
 & +\frac{1}{2}H_{\omega\omega}^{\prime\prime}\left(\omega-\omega_{e}\right)^{2}+...=0.\nonumber 
\end{align}
Since $H(k_{e},\omega_{e})=H_{k}^{\prime}(k_{e},\omega_{e})=0$, and
discarding the higher-order terms, 
\begin{equation}
k-k_{e}\simeq\pm\alpha_{1}\left(\omega-\omega_{e}\right)^{1/2}+\alpha_{2}\left(\omega-\omega_{e}\right)\pm\alpha_{3}\left(\omega-\omega_{e}\right)^{3/2}+O((\omega-\omega_{e})^{2})
\end{equation}
where
\begin{equation}
\alpha_{1}=\sqrt{-2\frac{H_{\omega}^{\prime}}{H_{kk}^{\prime\prime}}},\ \ \alpha_{2}=-\frac{H_{\omega k}^{\prime\prime}}{H_{kk}^{\prime\prime}},\ \ \alpha_{3}=\frac{\alpha_{1}}{2}\frac{(H_{\omega k}^{\prime\prime})^{2}-H_{\omega\omega}^{\prime\prime}H_{kk}^{\prime\prime}}{-2H_{\omega}^{\prime}H_{kk}^{\prime\prime}}.
\end{equation}
The coefficient $\alpha_{1}$ is the same as (\ref{eq:alpha_1}),
and the higher-order coefficients are the same as given in \cite{AM}
retaining the same order of terms.

\subsection{EDPs leading to branch points in the complex-frequency plane\label{subsec:EDPs-as-branch}}

Regarding Statements 2 and 3, it is clear from several points of view
that $D=0$ defines a degeneracy in the eigenvalue plane, and a square-root-type
BP in the complex-frequency plane (since $\mathbf{A}=\mathbf{A}\left(\omega\right)$
for $\mathbf{A=-\underline{\underline{\mathbf{Z}}}\,\underline{\underline{\mathbf{Y}}}}$,
$-\mathbf{\underline{\underline{\mathbf{Y}}}\,\underline{\underline{\mathbf{Z}}}}$,
or $\mathbf{\underline{\mathbf{M}}}$). Solving $D=0$ leads to the
frequency where the BP/EPD occurs (this also can be obtained by substituting
$k^{2}=-T/2$ from (\ref{tr}) into (\ref{det3})). Assuming for simplicity
that $\mathbf{\underline{\underline{\mathbf{G}}}}=\mathbf{\underline{\underline{\mathbf{0}}}}$,
this leads to 
\begin{equation}
\omega_{e}^{2}a+\omega_{e}b+c=0,
\end{equation}
where, 
for $L_{11}=L_{22}=L$ and $C_{11}=C_{22}=C$ ($C_{nm}$ is the $nm$th
element of the capacitance matrix), 
\begin{align}
a & =4\left(C_{12}L+CL_{12}\right)^{2},\\
b & =-4iC_{12}\left(CL_{12}+LC_{12}\right)\left(R_{11}+R_{22}\right),\nonumber \\
c & =-2R_{22}R_{11}\left(2C_{12}^{2}-C^{2}\right)-C^{2}\left(R_{11}^{2}+R_{22}^{2}\right).\nonumber 
\end{align}
If $\left(R_{11}+R_{22}\right)\neq0$, then $\omega_{e}$ will not
be on the real-$\omega$ axis, assuming $\left(CL_{12}+LC_{12}\right)\neq0$.

For the PT-symmetric case, $R_{11}=-R=-R_{22}$, 
\begin{equation}
\omega_{e}=\sqrt{\frac{-c}{a}}=R\frac{\sqrt{C^{2}-C_{12}^{2}}}{C_{12}L+CL_{12}}.\label{wBP}
\end{equation}
This will occur on the real-$\omega\,$\ axis, since one expects
$C^{2}>C_{12}^{2}$, proving Statement 4. Note that from a design
point of view, expression (\ref{wBP}) leads to the needed value of
$R$ for a desired value of $\omega_{e}$.

If we assume that $\mathbf{\underline{\underline{\mathbf{G}}}}\ne\mathbf{\underline{\underline{\mathbf{0}}}}$,
for $L_{11}=L_{22}=L$, $C_{11}=C_{22}=C$, and the PT-symmetric case,
$R_{11}=-R=-R_{22}$ and $G_{11}=-G=-G_{22}$, 
\[
\omega_{e}^{2}=-\frac{G^{2}L_{12}^{2}+R^{2}C_{12}^{2}-X-\left(G^{2}L^{2}+C^{2}R^{2}\right)}{\left(CL_{12}+LC_{12}\right)^{2}}
\]
where $X=2GR\left(LC+C_{12}L_{12}\right)$. If $R=0$, 
\begin{equation}
\omega_{e}=\frac{G\sqrt{L^{2}-L_{12}^{2}}}{C_{12}L+CL_{12}}
\end{equation}
which is the dual of (\ref{wBP}).

\begin{figure}[htp!]
{[}a{]}\includegraphics[width=0.4\textwidth]{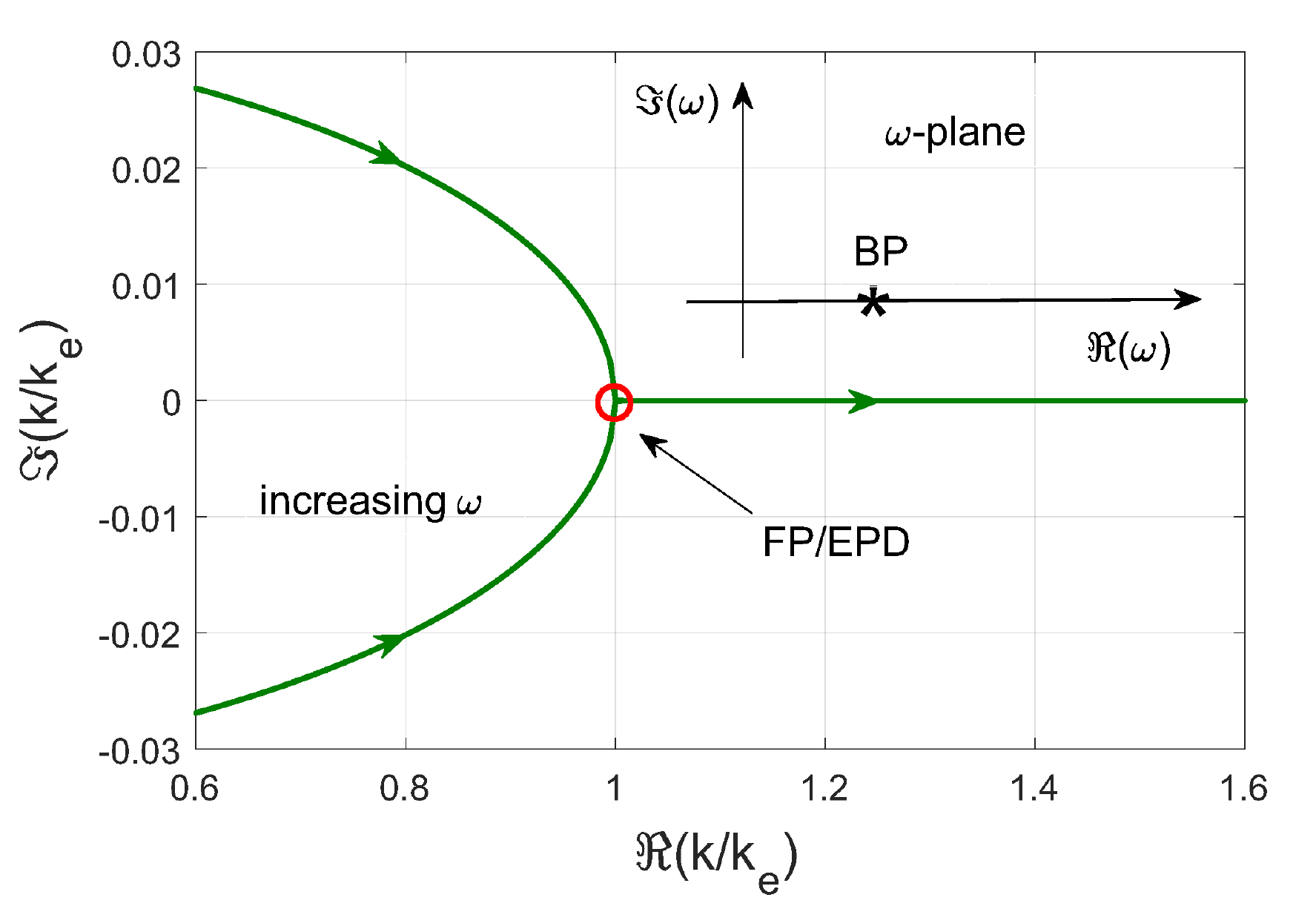}

{[}b{]}\includegraphics[width=0.4\textwidth]{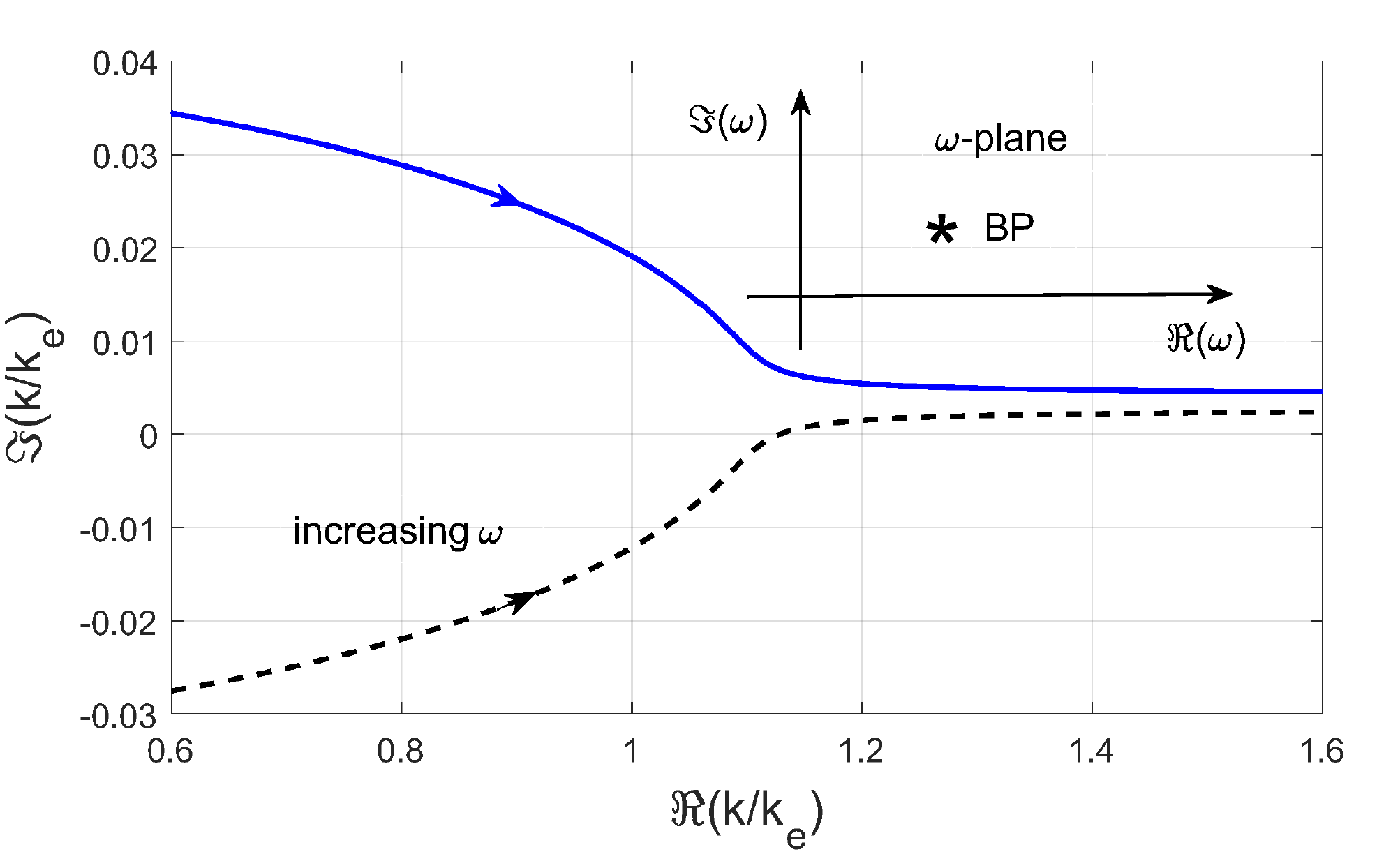}

{[}c{]}\includegraphics[width=0.4\textwidth]{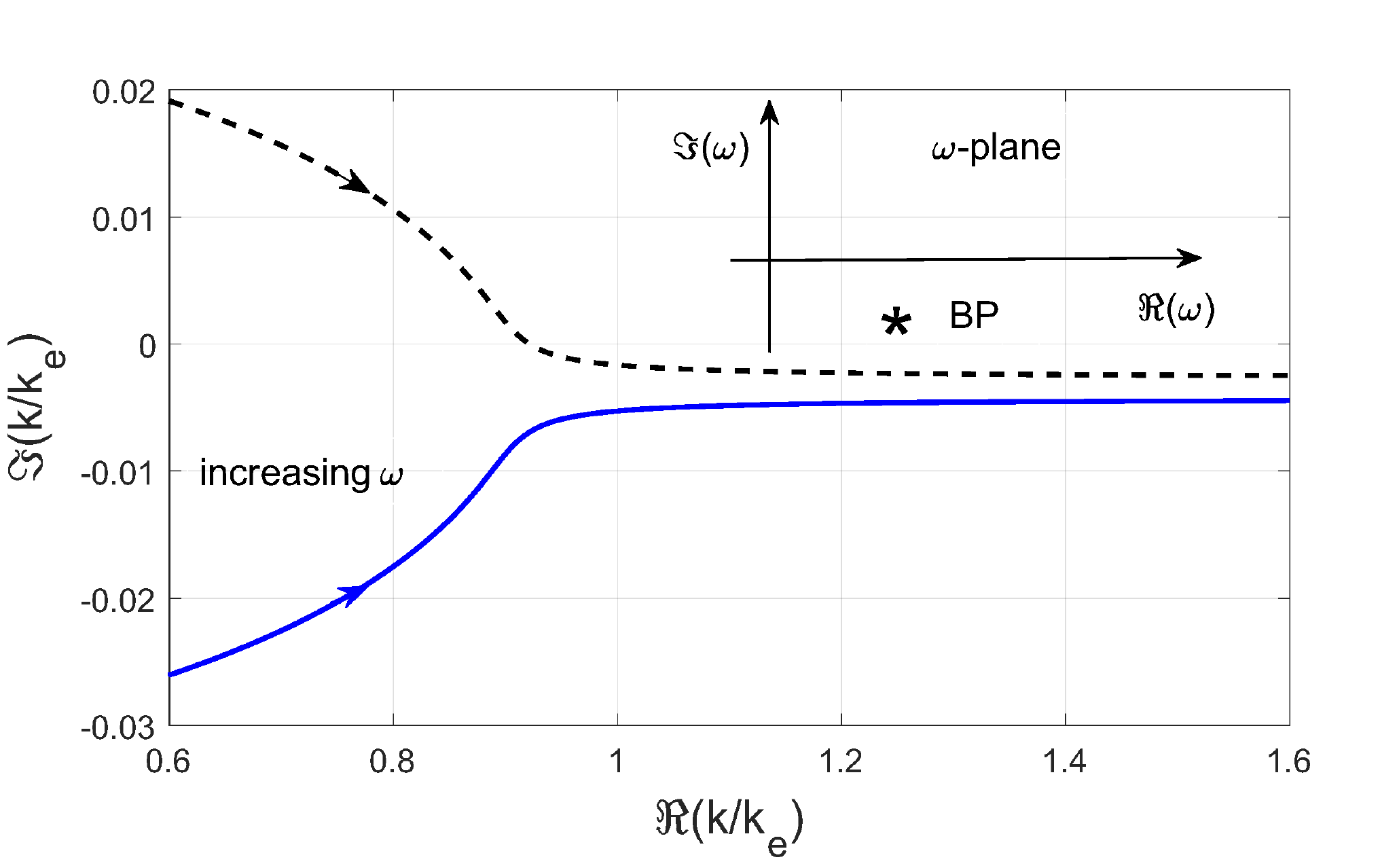}

{[}d{]}\includegraphics[width=0.4\textwidth]{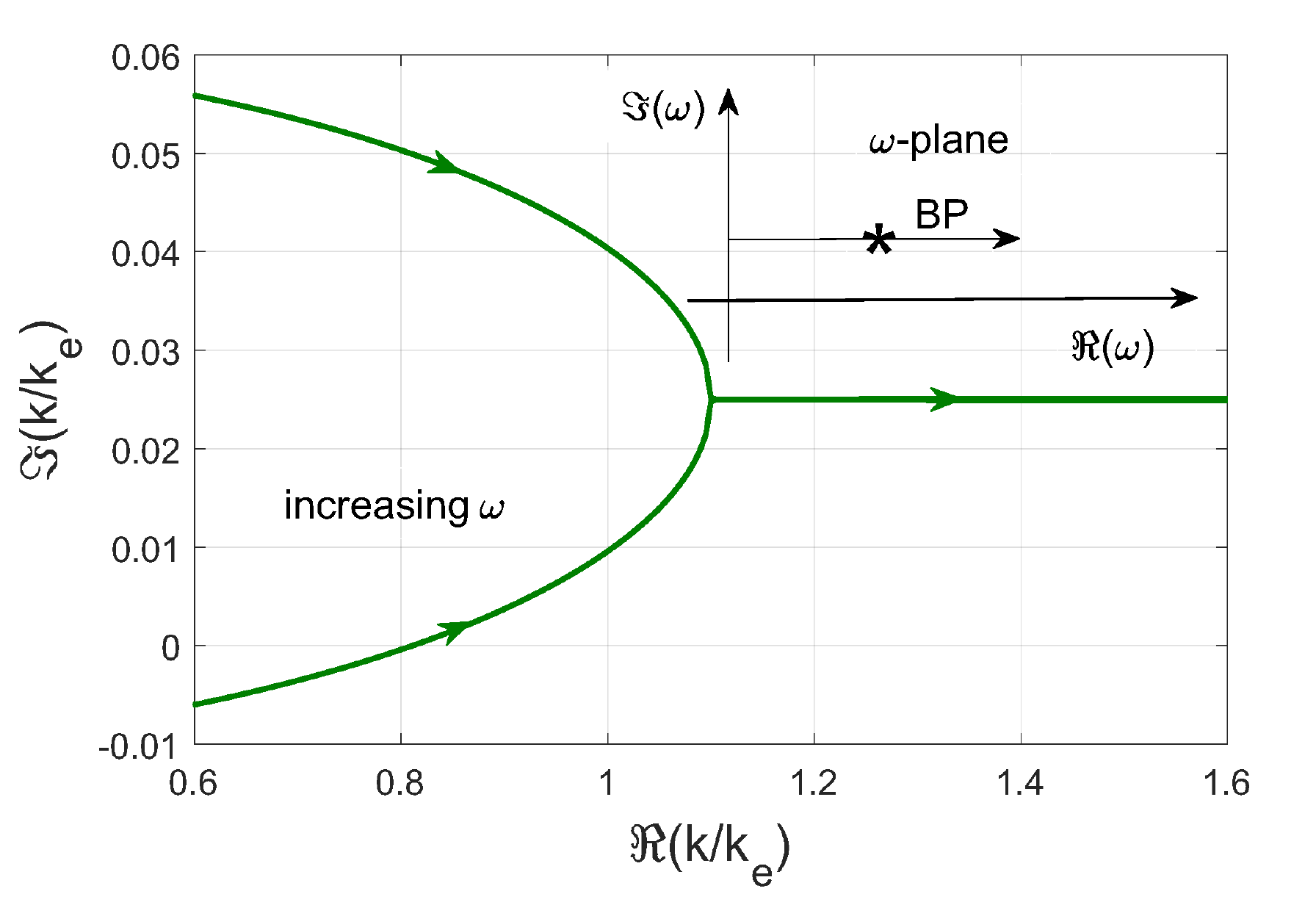}

\caption{Dispersion behavior near an EPD for coupled transmission lines with
(a) $R_{11}=-R_{22}=-73.172$ ohms (PT-symmetric case), as $\omega$
varies from $0.5\omega_{e}$ to $1.5\omega_{e}$ along the real-$\omega$
axis. (b) Same as (a) but for $R_{11}=-1.2R_{22}$, where the EPD
lies above the real-frequency axis. (c) Same as (a) but for $R_{11}=-0.8R_{22}$,
such that the EPD is below the real-frequency axis. (d) Same as (b)
but for $\Re(\omega)$ varying from $0.5\omega_{e}$ to $1.5\omega_{e}$
at a constant value $\Im(\omega)=\Im(\omega_{s})=(0.022\omega_{e})$.
In all cases the pair $(k_{e},\omega_{e})$ are the values at PT-symmetry,
$(k_{e},\omega_{e})=(28.649\,\mathrm{m^{-1}},2\pi10^{9}\,\mathrm{s^{-1}})$,
$R_{22}=73.172$ ohms, and the star indicates the BP/EPD. }
\label{fig2} 
\end{figure}
As an example, Fig. \ref{fig2} shows $\Im(k/k_{e})$ versus $\Re(k/k_{e})$
in the vicinity of the fold point for numerical parameters taken from
\cite{EPD} corresponding to two coupled microstrip lines (strip width
3 mm, gap between strips 0.1 mm; substrate height 0.75 mm, and dielectric
constant of 2.2); $C_{11}=C_{22}=C=0.12\,\,\textrm{nF/m},L_{11}=L_{22}=L=0.18\,\,\mu\textrm{H/m},L_{12}=L_{21}=49.24\,\,\textrm{nH/m},C_{12}=C_{21}=-25.83\,\,\textrm{pF/m},\textrm{ and }\mathbf{\underline{\underline{\mathbf{G}}}}=\mathbf{\underline{\underline{\mathbf{0}}}}$.
Setting a target frequency of $\omega_{e}=2\pi10^{9}\,\mathrm{s^{-1}}$,
from (\ref{wBP}), to place the EPD on the real frequency axis at
$\omega_{e}$ requires $R_{11}=-R_{22}=-73.172$ ohms. The corresponding
value of wavenumber at the EPD is $k_{e}=28.649$ ${\rm m}^{-1}$.
A two-dimensional root search of (\ref{Eq4-1})-(\ref{Eq4-1a}) yields
$(k_{s}/k_{e},\omega_{s}/\omega_{e})=(1,1)$ as expected. Dispersion
behavior in the vicinity of the fold point is shown in Fig. \ref{fig2}a.

For other values of $R_{11}=-R_{22}$ (i.e., maintaining PT-symmetry)
the fold point remains on the $\Re(\omega)$ axis, but moves to lower
or higher frequencies as indicated in (\ref{wBP}). Upon breaking
PT-symmetry by using $R_{11}\neq-R_{22}$, the BP/EPD does not occur
on the real-frequency axis, as shown in Figs. \ref{fig2}b,c,d, where
in all cases $R_{22}=73.172$ ohms. For $R_{11}=-1.2R_{22}$ the 2D
root search of (\ref{Eq4-1})-(\ref{Eq4-1a}) yields $(k_{s}/k_{e},\omega_{s}/\omega_{e})=(1.1+i0.025,1.1+i0.022)$,
where $(k_{e},\omega_{e})$ are the values given above under the PT-symmetry
conditions, $(k_{e},\omega_{e})=(28.649\,\mathrm{m^{-1}},2\pi10^{9}\,\mathrm{s^{-1}})$.
As such, the EPD lies above the real-frequency axis, and Fig. \ref{fig2}b
shows the corresponding dispersion behavior. Since a scanning of an
operating frequency (assumed real) does not pass through the branch
point, the eigenvalues do not become degenerate. Alternatively, Fig.
\ref{fig2}c shows the dispersion behavior when $R_{11}=-0.8R_{22}$,
such that the EPD is below the real-frequency axis and the modes have
interchanged with their counterparts in Fig. \ref{fig2}b. Fig. \ref{fig2}d
shows the dispersion behavior for the case $R_{11}=-1.2R_{22}$, when
the real part of frequency is varied while keeping a constant $\Im(\omega)=\Im(\omega_{s})=(0.022\omega_{e})$,
and so passing through the singular point (EPD), at which point the
modal degeneracy is recovered at a complex-valued $k$ . In this complex
frequency case a BP is clearly visible and occurs at a complex value
wavenumber. Regarding Figs. \ref{fig2}b,c, note that to interchange
the modal solutions it is not necessarily to encircle the EPD/BP (as
done in, for example \cite{Hassan}, \cite{R1}). It is shown in Figs.
\ref{fig2}b,c that the interchange of solutions is due to varying
the frequency path above or below the BP \cite{A6,A14}. 

\section{Conclusions}

We have examined several aspects of EPDs on two coupled transmission
lines, demonstrating that in the framework of the eigenvalue problem
the eigenvalue degeneracies are always coincident with eigenvector
degeneracies, such that all eigenvalue degeneracies correspond to
EPDs. We also discussed the fact that EPDs are related to branch-point
singularities in the complex-frequency plane, as can be ascertained
from both linear algebra concepts and from the theory of singular
points of complex mappings and bifurcation theory. Moreover, we have
provided a connection between the linear algebra approach and an approach
based on singularity and bifurcation theories, previously used to
study modal interactions on guided-wave structures. We have presented
simple closed-form expressions for the complex-frequency plane EPDs,
and showed that under PT-symmetry these branch points reside on the
real-frequency axis and generalized the branch point discussion to
complex frequency and wavenumbers. \bigskip{}

\section*{Acknowledgments}

This material is based upon M.O and F.C work supported by the Air
Force Office of Scientific Research under award number FA9550-15-1-0280. 

\bigskip{}

\section*{Appendix: Coupled-Mode Theory}

In addition to the transmission-line treatment of EPDs, here we briefly
comment on the matrix that arises from conventional so called ``coupled-mode
theory'' \cite{CMT}. For simplicity, we consider the PT-symmetric
case for otherwise identical individual transmission lines (e.g.,
one will have loss and one will have gain). Then, the individual (uncoupled)
lines have propagation constants $\beta$ and $\beta^{\ast}$, which,
when brought into proximity, become $\beta+\delta$ and $\beta^{\ast}+\delta^{\ast}$
under the coupling constant $\kappa$. The coupled system modes obey
the evolution equation \cite{PRL} 
\begin{equation}
i\frac{d}{dz}\left[\begin{array}{c}
a_{1}\\
a_{2}
\end{array}\right]=\left[\begin{array}{cc}
\beta+\delta & \kappa\\
\kappa^{\ast} & \left(\beta+\delta\right)^{\ast}
\end{array}\right]\left[\begin{array}{c}
a_{1}\\
a_{2}
\end{array}\right]=\mathbf{\underline{\underline{\beta}}}\left[\begin{array}{c}
a_{1}\\
a_{2}
\end{array}\right]\label{ee}
\end{equation}
where $a_{1}$ and $a_{2}$ are the wave amplitudes in transmission
lines 1 and 2, respectively. One can proceed with examination of the
eigenvectors and eigenvalues, but it suffices to consider, analogous
to (\ref{eq:H}), the dispersion relation 
\begin{eqnarray}
H\left(k,\omega\right) & = & \left\vert \left[\begin{array}{cc}
\beta\left(\omega\right)+\delta\left(\omega\right) & \kappa\left(\omega\right)\\
\kappa^{\ast}\left(\omega\right) & \left(\beta\left(\omega\right)+\delta\left(\omega\right)\right)^{\ast}
\end{array}\right]-k\left[\begin{array}{cc}
1 & 0\\
0 & 1
\end{array}\right]\right\vert =0\nonumber \\
 & = & k^{2}-k\mathrm{Tr}\left(\mathbf{\underline{\underline{\beta}}}\right)+\det\left(\mathbf{\mathbf{\underline{\underline{\beta}}}}\right)=0\label{ee1}
\end{eqnarray}
where $\mathbf{\mathbf{\underline{\underline{\beta}}}}$ is the 2$\times$2
matrix in (\ref{ee}). Obviously, (\ref{ee1}) is analogous to (\ref{HE0}).
Furthermore, 
\begin{equation}
H'_{k}\left(k,\omega\right)=2k-\left(\beta^{\ast}+\delta^{\ast}+\beta+\delta\right)=0
\end{equation}
leads to 
\begin{equation}
k=\frac{1}{2}\left(\beta^{\ast}+\delta^{\ast}+\beta+\delta\right)={\textstyle \Re}\left(\beta+\delta\right)=\frac{1}{2}\mathrm{Tr}\left(\mathbf{\mathbf{\mathbf{\underline{\underline{\beta}}}}}\right)
\end{equation}
and using (\ref{ee1}) one obtains 
\begin{equation}
\mathrm{Tr}^{2}\left(\underline{\underline{\beta}}\left(\omega\right)\right)-4\det\left(\underline{\underline{\beta}}\left(\omega\right)\right)=0
\end{equation}
which is the condition $D=0$ in (\ref{D}), and which leads to the
value of the EPD frequency $\omega=\omega_{\text{e}}$. The nonzero
condition $H_{\omega}^{\prime}(k,\omega)\neq0$ can be evaluated if
all matrix entries are known as a function of frequency. Thus, coupled-mode
theory leads to the same analysis of EPDs as the CTL formulation presented
in Section \ref{sec:Coupled-Transmission-Line}, and, therefore, can
also be analyzed using bifurcation theory.

\bigskip{}

\end{document}